\documentstyle[12pt]{article}

\renewcommand{\theequation}{\arabic{section}.\arabic{equation}}

\begin{document}

\begin{titlepage}

\rightline{BUHEP-92-33}
\rightline{October, 1992}

\begin{center}
{\LARGE\bf{Basis States for Hamiltonian QCD \\
                   with Dynamical Quarks}}

\bigskip
\bigskip
\bigskip

{\large Timothy E. Vaughan}

\bigskip
\bigskip

Department of Physics \\
Boston University \\
590 Commonwealth Ave. \\
Boston, MA 02215 \\
email: tvaughan@buphyk.bu.edu

\end{center}

\bigskip

\abstract{
\noindent We discuss the construction of basis states for Hamiltonian
QCD on the lattice, in particular states with dynamical quark pairs.  We
calculate the matrix elements of the operators in the QCD Hamiltonian
between these states.  Along with the ``harmonic oscillator'' states
introduced in previous pure SU(3) work, these states form a working
basis for calculations in full QCD.}

\end{titlepage}

\section{Introduction}
\setcounter{equation}{0}
As detailed in papers with Bronzan \cite{vaughan91a,vaughan91c}, we have
been studying lattice QCD using an
operator and states approach rather than the usual Monte Carlo
simulation of the Feynman path integral.  As we have emphasized, this
requires the use of a suitable set of single degree-of-freedom (DOF)
states on the SU(3) manifold.  Here ``suitable'' means that matrix
elements of the QCD Hamiltonian must be calculable in closed form using
these states, and that the states have tunable parameters so they
can model the QCD wave functions at all values of the coupling constant.

We have introduced a ``harmonic oscillator'' basis of states
as one which satisfies the above criteria and is therefore useful in
making calculations in Hamiltonian QCD.  Since these states describe
only gauge degrees of freedom, they are actually designed for studying
only a pure SU(3) gauge theory.

In this paper we will introduce a similarly suitable set of
dynamical-quark basis states to complement the harmonic oscillator
states.  We will discuss how we construct a quark ground state at
arbritrary values of the coupling constant, create quark
excitations, and calculate matrix elements of the QCD Hamiltonian
between these states.  Together with the harmonic oscillator states,
then, we have a complete basis for the study of full QCD.

The layout of this paper is as follows.  In Section~2 we briefly
review our work on the construction of harmonic oscillator
states.  In Section~3 we describe the construction of the quark
vacuum as well as the states which include virtual quark pairs.  In
Section~4 we discuss the combination of the harmonic oscillator states
with the quark states to form a basis for simulating full QCD.  In
Section~5 we derive the matrix elements of the operators of the QCD
Hamiltonian between these states.  We conclude in Section~6 with a
discussion of the usefulness of these states.

\section{Review of pure gauge states}
\setcounter{equation}{0}
Each gauge degree of freedom (link variable on the lattice) can be
described by a wave function which is a ``harmonic oscillator'' state on
the SU(3) manifold.  The harmonic oscillator states are derived from a
``Gaussian'' state on the manifold.  This Gaussian can be written
\begin{equation}
\psi_{\alpha_1}(\alpha,t) =
\sum_{p,q}d(p,q)e^{-t\lambda(p,q)}\chi^{(p,q)},
\end{equation}
where $d(p,q) = (p+1)(q+1)(p+q+2)/2$ is the dimension of the $(p,q)$
representation of SU(3), $\lambda(p,q) = (p+q)^2/4 + (p-q)^2/12 + (p+q)$
is the quadratic Casimir eigenvalue, and $\chi^{(p,q)}$ is the
character.  $\alpha \equiv (\alpha^1,\alpha^2,\dots,\alpha^8)$ is a
parameterization of the adjoint representation of SU(3).  The parameter
$t$ controls the width of the Gaussian, which is centered at $\alpha_1$.

To make connection with a more familiar object, note that the
corresponding wave function on the flat, three-dimensional manifold ${\bf
R}^3$ would be
\begin{equation}
\psi_{{\bf x}_1}({\bf x},t) = {{e^{-({\bf x}-{\bf x}_1)^2/4t}}\over
{(4\pi t)^4}},
\end{equation}
where ${\bf x}$ and ${\bf x}_1$ are ordinary vectors.  On this flat
manifold, harmonic oscillator states can be generated by applying a
polynomial in the operators $-i{\partial\over{\partial{x_{1_a}}}}$ to
the wave function, then setting ${\bf x}_1$ to 0.  On the SU(3)
manifold, the corresponding operators are ${\cal J}_{L_a}(\alpha_1)$, the
generators of SU(3) in differential form.\footnote{When applied to a
representation of SU(3), ${\cal J}_{L_a}(\alpha_1)$ has the effect of
left-multiplying the representation by the $a^{\em th}$ generator.
There are also operators ${\cal J}_{R_a}(\alpha_1)$ which right-multiply
the representation and could be used to generate a slightly different
set of harmonic oscillator states.}

The harmonic oscillator states that we have formed on the manifold
can be labeled by the number of SU(3) indices on the state.  Thus, the
zero-, one-, and two-color states are
\begin{eqnarray}
\phi = \phi({\bf\alpha},t) & = & \psi_{{\bf\alpha}_1}({\bf\alpha},t)
|_{{\bf\alpha}_1=0}, \nonumber\\
\phi_a = \phi_a({\bf\alpha},t) & = & {\cal J}_{L_a}({\alpha_1})
\psi_{{\bf\alpha}_1}({\bf\alpha},t)|_{{\bf\alpha}_1=0}, \nonumber \\
\phi_{ab} = \phi_{ab}({\bf\alpha},t) & = & \{{\cal
J}_{L_a}({\alpha_1}),{\cal J}_{L_b}
({\alpha_1})\}\psi_{{\bf\alpha}_1}({\bf\alpha},t)|_{{\bf\alpha}_1=0}.
\end{eqnarray}
For comparison, the analogous states on a {\em flat} manifold are
\begin{eqnarray}
\phi({\bf x},t) & = & e^{-{\bf x}^2/4t}, \nonumber \\
\phi_a({\bf x},t) & = & x_ae^{-{\bf x}^2/4t}, \nonumber \\
\phi_{ab}({\bf x},t) & = & x_ax_be^{-{\bf x}^2/4t}.
\end{eqnarray}
In the papers we have cited we described the orthogonalization of these
states as well as the calculation of matrix elements of the QCD
Hamiltonian between them.  We will not need those details here, however.

These single-DOF wave functions must now be pieced together to form
multi-DOF states suitable for computations in Hamiltonian QCD.  Each
multi-DOF state will be a product of single-DOF states.  The most basic
state is one with no excitations.  Specifically, each degree of freedom
is a zero-color state:
\begin{equation}
|\Psi\!> = |\phi\phi\phi\phi\cdots\!>.
\end{equation}
The next simplest states are those with just one excitation:
\begin{equation}
|\Psi\!> = |\phi_a\phi\phi\phi\cdots\!>.
\end{equation}
However, these states are not SU(3) singlets, as evidenced by the color
index which is left hanging.  In order to maintain global SU(3)
invariance, we must sum over all indices.  Therefore, the next possible
states are
\begin{equation}
|\Psi\!> = \delta_{ab}|\phi_a\phi_b\phi\phi\cdots\!>,
\end{equation}
where $\delta_{ab}$ is a Kronecker $\delta$-function.  There are of course
similar states where the excitations lie on different degrees of
freedom.  We call these {\em one-pair} states.  In analogous fashion, we
can form two-pair states, three-pair states, and so on.

Arbitrarily intricate states can be constructed by using highly colored,
single-DOF states along with complicated coupling schemes.  For
instance, the following is a valid (that is, gauge-invariant) state:
\begin{equation}
|\Psi\!> =
\delta_{ab}Tr(\lambda_c\lambda_d\lambda_e)
|\phi_{ac}\phi_{bd}\phi\phi_e\cdots\!>,
\end{equation}
where the $\lambda_i$'s are Gell-Mann matrices.

The states described above provide the basis for a pure SU(3) gauge
theory calculation, and they will likewise provide the basis for the
gauge sector of our full QCD calculation.  In the next sections
we will describe how to form similar states that include dynamical
quarks, to complete our basis.

\section{Quark sector states}
\setcounter{equation}{0}
To formulate our quark states, we use staggered fermions
\cite{Susskind}.  (We follow closely the notation of Banks
{\em et al.} \cite{banks}).  Specifically, the quarks are created and
destroyed by applications of single-component fields $\chi^\dagger_i({\bf
s})$ and $\chi_i({\bf s})$, where the index $i$ signifies color in the
fundamental representation of SU(3), and the argument ${\bf s} =
(x,y,z)$ specifies a site on the lattice.  The operators act on a simple
Fock space, and any quark configuration can be specified by giving the
location of the quarks on the lattice.

\subsection{Strong-coupling quark vacuum}
Just as we did for the gauge states, we would like to use quark states
which are flexible enough to treat both the strong- and weak-coupling
regimes.  Toward this end, we will begin with a description of the
strong-coupling vacuum, then discuss how to obtain the weak-coupling
vacuum from it.  We will then describe the excitations of the vacuum.

At very strong coupling the vacuum state is the ``checkerboard'' state,
in which alternating sites are fully occupied \cite{banks}.  Two of
these states can be formed, depending on which sublattice is filled (the
``red'' squares or the ``black'' squares of the checkerboard).  The two
states are
\begin{eqnarray}
|\psi_{even}\!> & = & \prod_{{\rm even~sites}}{1\over 6}\epsilon_{ijk}
\chi_i^\dagger({\bf s})\chi_j^\dagger({\bf s})\chi_k^\dagger({\bf s})
|E\!>, \nonumber \\
|\psi_{odd}\!>  & = & \prod_{{\rm odd~sites}}{1\over 6}\epsilon_{ijk}
\chi_i^\dagger({\bf s})\chi_j^\dagger({\bf s})\chi_k^\dagger({\bf s})
|E\!>,
\end{eqnarray}
where $|E\!>$ is the empty lattice, and $even/odd$ means
that $x+y+z = \pm 1$.\footnote{One must be careful to maintain a fixed
ordering for the product over the sites, lest minus sign errors creep
in.  Although we will not display it explicitly, we assume that the
sites have been ordered in some appropriate manner.}  These
two states transform into one another under a chirality transformation.
We can form two degenerate vacua, the symmetric and antisymmetric
combinations of these two states, which are eigenstates of the chirality
operator.  We have shown numerically that as we decrease the coupling
constant, the true vacuum is the symmetric combination, so we use
\begin{equation}
|0\!>_{strong} = |\psi_{even}\!> + \,|\psi_{odd}\!>
\end{equation}
as the vacuum state at strong coupling.

\subsection{Weak-coupling quark vacuum}
We derive the weak-coupling vacuum from the strong-coupling vacuum
by means of a projection operator.  We write
\begin{equation}
|0\!>_{weak} = \lim_{\alpha \rightarrow \infty} e^{-\alpha
H_w}|0\!>_{strong}.
\end{equation}
$H_w$ is the weak-coupling limit of the quark Hamiltonian,
\begin{equation}
H_w = {1\over{2a}}\sum_{{\bf s},{\bf s}'}\chi^{\dag}({\bf
s})M_{{\bf s}{\bf s}'}\chi({\bf s}'),
\end{equation}
where
\begin{equation}
M_{{\bf s}{\bf s}'} = \sum_{{\bf\hat n}}\bigl[\delta_{{\bf
s},{\bf s}'+{\bf\hat n}} + \delta_{{\bf s},{\bf
s}'-{\bf\hat n}}\bigr]\eta(\hat{\bf n}),
\end{equation}
$\eta(x) = (-1)^{z}$, $\eta(y) = (-1)^{x}$, $\eta(z) = (-1)^{y}$,
and $a$ is the lattice spacing.  (There is a suppressed color index in
this expression.)  Note that in weak coupling the quarks decouple from
the gauge degrees of freedom, and $H_w$ is equal to the free-field quark
Hamiltonian.

The projection works as follows.  $\alpha$ is an adjustable parameter
which governs the strength of the projection.  When $\alpha = 0$, there
is no projection, and we still have the strong-coupling vacuum.
As $\alpha$ becomes large, the projection operator damps out
the pieces of $|0\!>_{strong}$ which have high energy in the weak-coupling
limit, leaving intact the lowest energy weak-coupling state.  So, for
large $\alpha$, this operator does exactly what we want: it projects out
the weak-coupling vacuum.\footnote{Of course, we are assuming here that
the strong- and weak-coupling vacua are not orthogonal to each other.
This can be demonstrated {\em a posteriori}.}  In practice, $\alpha = 1$
will give nearly complete projection.  As we vary $\alpha$ from 0 to 1,
we obtain a vacuum state which interpolates smoothly between the
$g \rightarrow \infty$ and $g = 0$ vacua.  We therefore hope to obtain a
good approximate quark vacuum for arbitrary values of the coupling
constant by using $\alpha$ as a variational parameter in our
simulations.  We will designate this approximate vacuum $|0_q\!>$.

\subsection{Quark excitations}

Having established a quark-sector vacuum, we must now
understand how to create excited quark states.  This is much easier
and more familiar than it was for the gluon states, where we were
dealing with the complicated SU(3) manifold.  Here, we have a simple
Fock space, and we merely need to decide which states from that Fock
space to include in our basis.

We can exclude a large number of states by observing that the
``checkerboard'' vacuum state is half filled and that the Hamiltonian
preserves quark number.  We are interested in studying the glueball
spectrum, which has the same quantum numbers as the vacuum, so there is
no reason to include any quark state that isn't half-filled.  If
we wish to study the spectrum of particles with other quantum numbers,
we likewise limit ourselves to those sectors.

We can obtain a new half-filled state directly from the vacuum
state by applying one annihilation operator and one creation operator:
\begin{equation}
|\Psi\!> = \chi_{i}^{\dagger}({\bf s})\chi_{i}({\bf s'})|0_q\!>.
\end{equation}
Recall that the index $i$ indicates color, in the fundamental
representation of SU(3), reflecting the correct group properties of the
quarks.  Notice that we are summing over $i$, thereby maintaining
global gauge invariance in the quark sector in an analogous way to the
gauge sector.

The state shown above corresponds to exciting one virtual quark pair out
of the vacuum.  Clearly, we can form states with an arbitrary number of
virtual pairs by applying the appropriate number of creation and
annihilation operators.  One convenient feature of this scheme is that we
can control how many such pairs we wish to have in the simulation
simply by choosing which basis states to include.  We will comment on
this in the final section.

\section{Combining the gauge and quark states}
\setcounter{equation}{0}
Sections 2 and 3 described our construction of states for the gauge and
quark degrees of freedom in our simulations.  The simplest way to form a
combined state suitable for a full QCD simulation is to make a direct
product of one state from the gauge sector with one from the quark
sector.  A typical state, then could be written
\begin{equation}
|\Psi\!> =
\delta_{ab}\delta_{cd}|\phi_{a}\phi_{b}\phi_{c}\phi_{d}\phi\cdots\!>
\otimes\ \chi_{i}^{\dagger}({\bf s}_1)\chi_{i}({\bf s}_1')
\chi_{j}^{\dagger}({\bf s}_2)\chi_{j}({\bf s}_2')|0_q\!>.
\end{equation}
Let us emphasize again that all of the SU(3) indices are contracted,
ensuring that global gauge invariance is maintained.  Note in particular
that the indices are summed separately within the two sectors,
indicating that the gauge and quark sectors are separately invariant
under the global SU(3) transformation.

There is a second way to form combined states, in which the two sectors
are not separately invariant.  To do this, we need an object which can
intertwine the SU(3) adjoint representation of the gauge sector with the
fundamental representation of the quark sector.  The simplest way to do
this is with the Gell-Mann matrices, which carry both three- and
eight-dimensional indices.  A simple state of this form is
\begin{equation}
|\Psi\!> = |\phi_{a}\phi\phi\phi\phi\!> \otimes
\ \chi_{i}^{\dagger}({\bf s})\lambda^{a}_{ij}\chi_{j}({\bf
s'})|0_q\!>.
\end{equation}

This completes the description of our basis states for full QCD
calculations on the lattice.  In the next section we will show how to
compute the matrix elements necessary for our simulations.
\section{Matrix elements}
\setcounter{equation}{0}
The matrix elements of our combined states are sums of products of gauge
matrix elements with quark matrix elements.  The results of our gauge
sector calculations have already appeared in \cite{vaughan91c}, and we
will not repeat them here.

The most basic quark matrix element that we must calculate is the
overlap of our projected vacuum, $<\!0_q|0_q\!>$.  This, in turn, can be
broken down into sums of matrix elements of the projected checkerboard
states:
\begin{eqnarray}
<\!0_q|0_q\!> & = & <\psi_{even}|e^{-2\alpha H_w}|\psi_{even}\!>
                +   <\psi_{even}|e^{-2\alpha H_w}|\psi_{odd}\!>
\nonumber \\
              & + & <\psi_{odd}|e^{-2\alpha H_w}|\psi_{even}\!>
                +   <\psi_{odd}|e^{-2\alpha H_w}|\psi_{odd}\!>.
\end{eqnarray}
We will explicitly calculate only the first of these four nearly identical
expressions.  Recall that
\begin{eqnarray}
|\psi_{even}\!> & = & \prod_{{\rm even~sites}}{1\over 6}\epsilon_{ijk}
\chi_i^\dagger({\bf s})\chi_j^\dagger({\bf s})\chi_k^\dagger({\bf s})
|E\!> \nonumber \\
                & = & \bigl[\chi_1^\dagger({\bf s}_1)\chi_2^\dagger({\bf
s}_1)\chi_3^\dagger({\bf s}_1)\bigr]\cdots\bigl[\chi_1^\dagger({\bf
s}_M)\chi_2^\dagger({\bf
s}_M)\chi_3^\dagger({\bf s}_M)\bigr]|E\!>,
\end{eqnarray}
where $M={N^3\over 2}$ is the number of even sites on our $N\times
N\times N$ lattice,\footnote{We assume $N$ is an even number.} and ${\bf
s}_1,\ldots,{\bf s}_M $ specify the locations of those sites.
First, we segregate the creation operators by color.  This
involves an even number of permutations, so no minus sign is picked up.
\begin{equation}
|\psi_{even}\!> = \bigl[\chi_1^\dagger({\bf
s}_1)\cdots\chi_1^\dagger({\bf s}_M)\bigr]\bigl[\chi_2^\dagger({\bf
s}_1)\cdots\chi_2^\dagger({\bf s}_M)\bigr]\bigl[\chi_3^\dagger({\bf
s}_1)\cdots\chi_3^\dagger({\bf s}_M)\bigr]|E\!>.
\end{equation}
Likewise, we can segregate $H_w$ into a product of
single-color factors:
\begin{eqnarray}
H_w & \equiv & H_w^1 + H_w^2 + H_w^3 \nonumber \\
         & =      & {1\over{2a}}\sum_{{\bf s},{\bf
s}'}\bigl[\chi_1^{\dag}({\bf s})M_{{\bf s}{\bf
s}'}\chi_1({\bf s}') + \chi_2^{\dag}({\bf s})M_{{\bf s}{\bf
s}'}\chi_2({\bf s}') + \chi_3^{\dag}({\bf s})M_{{\bf s}{\bf
s}'}\chi_3({\bf s}')\bigr].~~~~~~~~
\end{eqnarray}
Since the separate color factors commute, we can write
\begin{equation}
e^{-2\alpha H_w} = e^{-2\alpha H_w^1}e^{-2\alpha
H_w^2}e^{-2\alpha H_w^3}
\end{equation}
and thereby write our matrix element in color-factorized form:
\begin{eqnarray}
<\psi_{even}|e^{-2\alpha H_w}|\psi_{even}\!> & = &
<\!E|\bigl[\chi_1({\bf s}_M)\cdots\chi_1({\bf s}_1)\bigr]e^{-2\alpha
H_w^1}\bigl[\chi_1^\dagger({\bf
s}_1)\cdots\chi_1^\dagger({\bf s}_M)\bigr] \nonumber \\
 & & ~~~~\times\bigl[\chi_2({\bf s}_M)\cdots\chi_2({\bf
s}_1)\bigr]e^{-2\alpha H_w^2}\bigl[\chi_2^\dagger({\bf
s}_1)\cdots\chi_2^\dagger({\bf s}_M)\bigr] \nonumber \\
 & & ~~~~\times\bigl[\chi_3({\bf s}_M)\cdots\chi_3({\bf
s}_1)\bigr]e^{-2\alpha H_w^3}\bigl[\chi_3^\dagger({\bf
s}_1)\cdots\chi_3^\dagger({\bf s}_M)\bigr]|E\!>. \nonumber \\
 & &
\end{eqnarray}

We now want to commute the exponential factors all the way to the left,
where we will be able to take advantage of the fact that
\begin{equation}
<\!E|e^{-2\alpha H_w^i} = <\!E|.
\end{equation}
To do this, we will need the relation
\begin{equation}
\chi({\bf s})e^{-\sigma H_w} = \sum_{{\bf
s}'}K_{{\bf s},{\bf s}'}(\sigma)e^{-\sigma
H_w}\chi({\bf s}'),
\end{equation}
where
\begin{equation}
K_{{\bf s},{\bf s}'}(\sigma) = \sum_{{\bf k},i}\phi({\bf
s};{\bf k},i)e^{\sigma\lambda({\bf k},i)}\phi({\bf
s}';{\bf k},i)
\end{equation}
and the functions $\phi({\bf s};{\bf k},i)$ are the eigenfunctions of
$M_{{\bf s}{\bf s}'}$.  We prove this relation in the Appendix.

Commuting each of the exponential factors to the left gives
\begin{eqnarray}
<\psi_{even}|e^{-2\alpha H_w}|\psi_{even}\!> & = &
<\!E|\prod_{i=1}^M\bigl[\sum_{{\bf s}_i'}K_{{\bf s}_i,{\bf
s}_i'}(2\alpha)\chi_1({\bf s}_i')\bigr]
\chi_1^\dagger({\bf s}_1)\cdots\chi_1^\dagger({\bf s}_M) \nonumber \\
 & & ~~~~\times\prod_{i=1}^M\bigl[\sum_{{\bf s}_i'}K_{{\bf
s}_i,{\bf s}_i'}(2\alpha)\chi_2({\bf
s}_i')\bigr]\chi_2^\dagger({\bf s}_1)\cdots\chi_2^\dagger({\bf
s}_M) \nonumber \\
 & & ~~~~\times\prod_{i=1}^M\bigl[\sum_{{\bf s}_i'}K_{{\bf
s}_i,{\bf s}_i'}(2\alpha)\chi_3({\bf
s}_i')\bigr]\chi_3^\dagger({\bf s}_1)\cdots\chi_3^\dagger({\bf
s}_M)|E\!>. \nonumber \\
 & &
\end{eqnarray}
The $M$ sites ${\bf s}_1',\ldots,{\bf s}_M'$ can be
chosen in any way from the even sites, but unless they are a permutation
of ${\bf s}_1,\ldots,{\bf s}_M$, the contribution to the matrix element
is zero.  Thus, we get
\begin{eqnarray}
<\psi_{even}|e^{-2\alpha H_w}|\psi_{even}\!> & = &
<\!E|\sum_{\rm perms~P}K_{{\bf s}_1,{\bf s}_{P(1)}}\cdots K_{{\bf
s}_M,{\bf s}_{P(M)}}\epsilon_{P(1)\cdots P(M)} \nonumber \\
 & &~~~~\times\bigl[\chi_1({\bf s}_1)\cdots\chi_1({\bf
s}_M)\bigr]\bigl[\chi_1^\dagger({\bf s}_1)\cdots\chi_1^\dagger({\bf
s}_M)\bigr] \nonumber \\
 & &~~~~\times\{{\rm Similar~factors~for~other~colors}\}|E\!> \nonumber \\
 & = & \bigl[\sum_{\rm perms~P}K_{{\bf s}_1,{\bf s}_{P(1)}}\cdots K_{{\bf
s}_M,{\bf s}_{P(M)}}\bigr]^3.
\end{eqnarray}
Introduce the $M\times M$ matrix $L^{{\rm EE}}(\alpha)$, whose
$(m,n)^{\rm th}$ element is given by
\begin{equation}
L_{mn}^{{\rm EE}}(\alpha) = K_{{\bf s}_m,{\bf s}_n}(2\alpha).
\end{equation}
The EE refers to the fact that we are calculating the even-even
matrix element.  We can then write our even-even matrix element in
its final, deceptively compact form:
\begin{equation}
<\psi_{even}|e^{-2\alpha H_w}|\psi_{even}\!> =
\bigl[\det L^{EE}(\alpha)\bigr]^3.
\end{equation}
The even-odd, odd-even, and odd-odd pieces of $<\!0_q|0_q\!>$ can be
calculated in analogous fashion, the only difference being the
composition of the corresponding $L$ matrix.  We therefore have
completed the calculation of the norm of the quark vacuum:
\begin{equation}
<\!0_q|0_q\!> = \bigl[\det L^{{\rm EE}}(\alpha)\bigr]^3 + \bigl[\det
L^{{\rm EO}}(\alpha)\bigr]^3 + \bigl[\det L^{{\rm OE}}(\alpha)\bigr]^3 +
\bigl[\det L^{{\rm OO}}(\alpha)\bigr]^3.
\end{equation}

Now that the machinery is set up, the remaining matrix elements require
much less effort.  The next simplest matrix element to calculate is
\begin{equation}
<\!0_q|\chi_i({\bf s}_0)\chi_j^\dagger({\bf s}_0')|0_q\!>.
\end{equation}
We need not consider operators $\chi\chi$ or $\chi^\dagger\chi^\dagger$,
which change quark number and therefore have trivially vanishing
matrix elements.  (As we have mentioned already, the QCD Hamiltonian
preserves quark number, and therefore does not include operators of
this type.)  Similarly, if $i\ne j$, then the matrix element vanishes
because there would be a mismatch in the number of quarks of a given
color.

The calculation of this matrix element proceeds exactly as the overlap
calculation, except that one of the colors (suppose it's color 1) will
have the extra operators $\chi_1({\bf s}_0)$ and $\chi_1^\dagger({\bf
s}_0')$ to commute through the exponential.  This gives color 1 a
contribution of
\begin{eqnarray}
 & & \sum_{\overline{{\bf s}}_0,\overline{{\bf s}}_0'}
K_{{\bf s}_0\overline{{\bf s}}_0}(-\alpha)K_{\overline{{\bf s}}_0'{\bf
s}_0'}(-\alpha)<\!E|\bigl[\chi_1({\bf s}_M)\cdots\chi_1({\bf
s}_1)\chi_1(\overline{{\bf s}}_0)\bigr] \nonumber \\
 & & ~~~~~~~~~~~~~~~~~~~~~~~~~~~~\times e^{-2\alpha
H_w^1}\bigl[\chi_1^\dagger(\overline{{\bf
s}}_0')\chi_1^\dagger({\bf s}_1)\cdots\chi_1^\dagger({\bf
s}_M)\bigr]|E\!>.
\end{eqnarray}
If we define the extended $(M+1)\times(M+1)$ matrix
\begin{equation}
L^{\rm EE}({\bf s}_0,{\bf s}_0') = \left(
\begin{array}{c|ccc}
        \delta_{{\bf s}_0,{\bf s}_0'} & K_{{\bf s}_0,{\bf s}_1}(\alpha)
& \cdots & K_{{\bf s}_0,{\bf s}_M}(\alpha) \\ \hline
        K_{{\bf s}_1,{\bf s}_0'}(\alpha) & K_{{\bf s}_1,{\bf s}_1}(2\alpha) &
\cdots
& K_{{\bf s}_1,{\bf s}_M}(2\alpha) \\
        \vdots & \vdots & \ddots & \vdots \\
        K_{{\bf s}_M,{\bf s}_0'}(\alpha) & K_{{\bf s}_M,{\bf s}_1}(2\alpha) &
\cdots
& K_{{\bf s}_M,{\bf s}_M}(2\alpha)
\end{array}
\right),
\end{equation}
then we can again write a compact expression for our matrix element,
\begin{eqnarray}
<\!0_q|\chi_i({\bf s}_0)\chi_j^\dagger({\bf s}_0')|0_q\!> & = & \delta_{ij}
\det L^{\rm EE}({\bf s}_0,{\bf s}_0')[\det L^{\rm EE}(\alpha)]^2 \nonumber \\
 & + & {\rm EO~+~OE~+~OO~contributions}.
\end{eqnarray}
The factors of $\det L^{\rm EE}(\alpha)$ are the contribution of the two
unmodified colors.

Matrix elements involving more $\chi\chi^\dagger$ pairs can be
calculated in analogy with the above derivation.  Each additional pair
requires the introduction of a further extended matrix, for the case in
which all the $\chi$'s act on the same color.  For instance, the next
extension of the matrix comes from
\begin{eqnarray}
<\!0_q|\chi_1({\bf s}_0)\chi_1({\bf s}_{00})\chi_1^\dagger({\bf
s}_0')\chi_1^\dagger({\bf s}_{00}')|0_q\!> & = &
\det L^{\rm EE}({\bf s}_0,{\bf s}_0';{\bf s}_{00},{\bf s}_{00}')[\det
L^{\rm EE}(\alpha)]^2 \nonumber \\
 & + & {\rm EO~+~OE~+~OO~contributions}, \nonumber \\
 & &
\end{eqnarray}
where
\begin{equation}
L^{\rm EE}({\bf s}_0,{\bf s}_0';{\bf s}_{00},{\bf s}_{00}') = \left(
\begin{array}{cc|ccc}
        \delta_{{\bf s}_{00},{\bf s}_{00}'} & \delta_{{\bf s}_{00},{\bf s}_0'}
&
K_{{\bf s}_{00},{\bf s}_1}(\alpha) & \cdots & K_{{\bf s}_{00},{\bf
s}_M}(\alpha) \\
        \delta_{{\bf s}_{00},{\bf s}_0'} & \delta_{{\bf s}_0,{\bf s}_0'} &
K_{{\bf s}_0,{\bf s}_1}(\alpha) & \cdots & K_{{\bf s}_0,{\bf
s}_M}(\alpha) \\ \hline
        K_{{\bf s}_1,{\bf s}_{00}'}(\alpha) & K_{{\bf s}_1,{\bf
s}_0'}(\alpha) & K_{{\bf s}_1,{\bf s}_1}(2\alpha) & \cdots
& K_{{\bf s}_1,{\bf s}_M}(2\alpha) \\
        \vdots & \vdots & \vdots & \ddots & \vdots \\
        K_{{\bf s}_M,{\bf s}_{00}'}(\alpha) & K_{{\bf s}_M,{\bf
s}_0'}(\alpha) & K_{{\bf s}_M,{\bf s}_1}(2\alpha) & \cdots
& K_{{\bf s}_M,{\bf s}_M}(2\alpha)
\end{array}
\right).
\end{equation}

We have calculated matrix elements of this type for up to four
$\chi\chi^\dagger$ pairs, which is the requirement for simulations with
two dynamical quark pairs.\footnote{If one fixes the gauge by our current
method\cite{vaughan92a}, the QCD Hamiltonian can contribute up to two
pairs of operators, rather than just one.  In this case we would need
two more ``extensions'' of the $L$ matrix to do the equivalent
simulations.}
\section{Summary}
\setcounter{equation}{0}
We have introduced a set of basis states that is suitable for making
calculations in an ``operator and states'' approach to the study of
lattice QCD.  The previous section detailed the computation of matrix
elements of the quark-sector states.  In previous work we calculated
matrix elements for pure gauge simulations.  Taken together, these
matrix elements provide us the tools to construct a Hamiltonian matrix
on a basis of states in full QCD.  The eigenvalues of this matrix are
then estimates of masses in the QCD spectrum.

We would like to emphasize the particular power that this method has
with respect to testing the quenched approximation.  Unlike Monte Carlo
simulations of the Feynman path integral, dynamical quarks pose no
particular problems to our method.  The states which include quark pairs
are ``just another configuration.''  Moreover, by choosing which states
to include, we can limit the simulation to having as many virtual quark
pairs as we wish.  By comparing runs with and without quark pairs, we
hope to be able to provide some insight into the quenched approximation.

We would like to acknowledge the contributions made to this work by
J.~B.~Bronzan.
\appendix

\section{Appendix}
\renewcommand{\theequation}{A.\arabic{equation}}
\setcounter{equation}{0}
To prove the relation
\begin{equation}
\chi({\bf s})e^{-\sigma H_w} = \sum_{{\bf
s}'}K_{{\bf s},{\bf s}'}(\sigma)e^{-\sigma
H_w}\chi({\bf s}'),
\end{equation}
we first need to diagonalize $H_w.$  Recall that $H_w$ is
bilinear in the quark creation and annihilation operators:
\begin{equation}
H_w = {1\over {2a}}\sum_{{\bf s},{\bf s}'}\chi^\dagger({\bf
s})M_{{\bf s}{\bf s}'}\chi({\bf s}'),
\end{equation}
where
\begin{equation}
M_{{\bf s}{\bf s}'} = \sum_{{\bf\hat n}}\bigl[\delta_{{\bf
s},{\bf s}'+{\bf\hat n}} + \delta_{{\bf s},{\bf
s}'-{\bf\hat n}}\bigr]\eta(\hat{\bf n}).
\end{equation}
Diagonalizing $M_{{\bf s}{\bf s}'}$ amounts to solving the Dirac
equation on the lattice.  We introduce quark mode operators, which
annihilate ``plane-wave'' states:
\begin{equation}
\chi({\bf k},i) = \sum_{\bf s}\chi({\bf s})\phi({\bf s};{\bf k},i).
\end{equation}
The $\phi({\bf s};{\bf k},i)$ are the eigenfunctions of $M_{{\bf s}{\bf
s}'}$:
\begin{equation}
\sum_{\bf s'}M_{{\bf s}{\bf s}'}\phi({\bf
s}';{\bf k},i) = \lambda({\bf k},i)\phi({\bf s};{\bf k},i),
\end{equation}
where ${\bf k}$ is the lattice momentum, $i$ enumerates the
modes, and $\lambda({\bf k},i)$ is the eigenvalue of the
mode.\footnote{The plane-wave operators also have a color index, which
we suppress here.}  On a lattice with $N\times N\times N$ sites, $1\le
k_x,k_y,k_z\le {N\over 2}$ and $1\le i\le N^3$.  The derivation of the
eigenfunctions is straightforward but rather tedious.  We have carried
it out explicitly only for a $2\times 2\times 2$ lattice.

The weak-coupling quark Hamiltonian is (by design) diagonal using these
new operators:
\begin{eqnarray}
H_w & = & {1\over {2a}}\sum_{{\bf s},{\bf s}'}\chi^\dagger({\bf
s})M_{{\bf s}{\bf s}'}\chi({\bf s}') \nonumber \\
         & = & {1\over {2a}}\sum_{{\bf s},{\bf s}'}\sum_{{\bf
k},{\bf k}'}\sum_{i,i'}\chi^\dagger({\bf k},i)\phi({\bf s};{\bf
k},i)M_{{\bf s}{\bf s}'}\phi({\bf s}';{\bf k}',i')\chi({\bf k}',i')
\nonumber \\
          & = & {1\over {2a}}\sum_{{\bf k},i}\lambda({\bf
k},i)\chi^\dagger({\bf k},i)\chi({\bf k},i).
\end{eqnarray}

Next, we define the expression
\begin{equation}
S(\sigma) = e^{\sigma H_w}\chi({\bf k},i)e^{-\sigma H_w}.
\end{equation}
Then
\begin{equation}
{dS(\sigma)\over d\sigma} = e^{\sigma H_w}[H_w,\chi({\bf
k},i)]e^{-\sigma H_w}.
\end{equation}
Substituting the diagonal form of $H_w$ we just derived, we have
\begin{equation}
[H_w,\chi({\bf k},i)] = -\lambda({\bf k},i)\chi({\bf k},i).
\end{equation}
Therefore,
\begin{equation}
{dS(\sigma)\over d\sigma} = -\lambda({\bf k},i)S(\sigma),
\end{equation}
which implies that
\begin{equation}
S(\sigma) = {\rm constant}\times e^{-\sigma\lambda({\bf k},i)}.
\end{equation}
Noting that $S(0) = \chi({\bf k},i)$, we have
\begin{equation}
S(\sigma) = e^{-\sigma\lambda({\bf k},i)}\chi({\bf k},i),
\end{equation}
giving us the relation
\begin{equation}
e^{\sigma H_w}\chi({\bf k},i)e^{-\sigma H_w} = e^{-\sigma\lambda({\bf
k},i)}\chi({\bf k},i).
\end{equation}
If we now multiply both sides of this by $\phi({\bf s};{\bf k},i)$,
sum over ${\bf k}$ and $i$, and change back to configuration space
variables $\chi({\bf s})$, we obtain
\begin{equation}
e^{\sigma H_w}\chi({\bf s})e^{-\sigma H_w} = \sum_{{\bf s}'}\sum_{{\bf
k},i}\phi({\bf s};{\bf k},i)e^{-\sigma\lambda({\bf k},i)}\phi({\bf
s}';{\bf k},i)\chi({\bf s}').
\end{equation}
Substituting the variable $K$, and left-multiplying by $e^{-\sigma
H_w}$, we get the relation we set out to prove,
\begin{equation}
\chi({\bf s})e^{-\sigma H_w} = \sum_{{\bf
s}'}K_{{\bf s},{\bf s}'}(\sigma)e^{-\sigma
H_w}\chi({\bf s}').
\end{equation}

\bibliographystyle{plain}

\end{document}